\documentclass{mpe_report}

\usepackage{psfig,graphicx,epsfig}
\usepackage{color}
\usepackage{amsmath,amssymb,epic,eepic,array}

\begin{document}

\pagenumbering{arabic}
\setcounter{page}{56}

\renewcommand{\FirstPageOfPaper }{ 56}\renewcommand{\LastPageOfPaper }{ 59}

\title{The 8gr8 Cygnus Survey for New Pulsars and RRATs}
\author{E. Rubio-Herrera\inst{1}, R. Braun\inst{2}, R.T. Edwards\inst{3}, G. Janssen\inst{1}, J. van Leeuwen\inst{4}, B.W. Stappers\inst{1,2}, W. van Straten\inst{5}}  
\institute{Astronomical Institute \emph{Anton Pannekoek}, University of Amsterdam,
Kruislaan 403, 1098 SJ Amsterdam The Netherlands.
\and  ASTRON, P.O. Box 2 -- 7990 AA Dwingeloo, The Netherlands
\and  Australia Telescope National Facility, CSIRO, PO Box 76, Epping, NSW 1710, Australia
\and  UBC Astronomy, Canada UBC Astronomy, 6224 Agricultural Rd., Vancouver B.C. V6T 1Z1, Canada
\and  Center for Gravitational Wave Astronomy, The University of Texas at Brownsville 80 Fort Brown Brownsville, TX 78520 U.S.A}
\authorrunning{E. Rubio--Herrera \emph{et. al.}}
\titlerunning{The 8gr8 Cygnus Survey}
\maketitle

\begin{abstract}
We are currently undertaking a survey to search for new pulsars and the recently found Rotating RAdio Transcients (RRATs) in the Cygnus OB complex.
The survey uses the Westerbrok Synthesis Radio Telescope (WSRT) in a unique way called the 8gr8 mode, which gives it the best efficiency of any low-frequency wide-area
survey. So far we have found a few new pulsars and the routines for the detection of RRATs are already implemented in the standard reduction. 
We expect to find a few tens of new pulsars and a similar number of RRATs. This will help us to improve our knowledge about the population and properties 
of the latter poorly known objects as well as provide an improved knowledge of the number of young pulsars associated with the OB complexes in the Cygnus region.
\end{abstract}

\section{The 8gr8 survey of pulsars and RRATs}

The 8gr8 Cygnus Survey will cover the region of the Galactic plane located between
$100^{\circ} < \ell < 40^{\circ}$ and $-0.25^{\circ} < b < 7.25^{\circ}$ respectively, covering around 420 sq deg.
in Cygnus. 
This region is known as the Cygnus superbubble~(\cite{UFRAW}), contains a lot of OB associations and  hot gas
that might be generated by SN explosions, producing a large amount of compact objects ~(\cite{PG}).

\begin{figure}
\centerline{\psfig{file=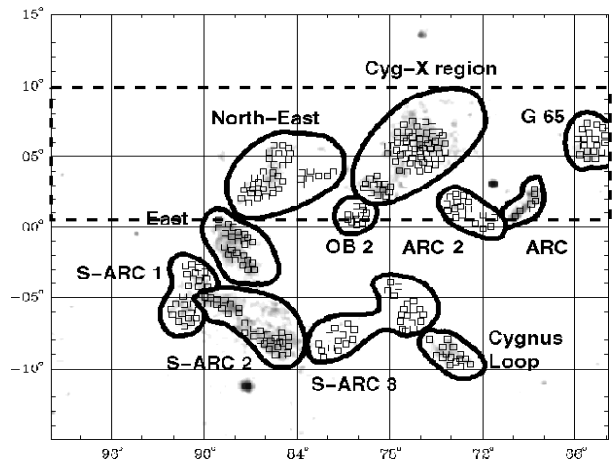,height=6.0cm,width=8.0cm,clip=} }
\caption{The Cygnus Superbubble. the dashed line box shows approximately the area covered by the 8gr8 survey. Figure taken from ~\cite{UFRAW}.
\label{image}}
\end{figure}

In these regions we expect to find significant numbers of young pulsars.
~\cite{DTWS} made a previous survey in this area finding a few tens of new pulsars.
In the 8gr8 survey, we expect to have about 5 times better sensitivity than the previous surveys, 
for pulsars with $P \geq 1$ s, and more than an order of magnitude for those with $P \leq 1$ s.
The 8gr8 survey will cover an area of the sky which has not been revisited in the latest big surveys for pulsars and according
to our simulations we expect to find a few tens of pulsars.

\subsection{Current pulsar surveys}
Recent pulsar surveys have not been able to cover all of this region of Galactic plane and this survey provides an excellent way to do this with
high sensitivity. 
To date the most successful survey  is the Parkes Multibeam Survey (\cite{Manch}) which has 
discovered approximately half of the known pulsars.

This survey covered a region of the sky located
 along the Galactic plane at low latitude ($|b| < 5^{\circ}$) and with a Galactic longitude of $32^{\circ} < \ell < 77 ^{\circ}$.
This survey applied a multibeam technique using a receiver with 13 beams.

Another remarkable survey is the Arecibo-P Alfa survey (\cite{Cord}). This survey expects to find
a few hundred new pulsars and is concentrated
 at very low Galactic latitudes $|b| < 5^{\circ}$ and along the region of the Galactic plane visible from Arecibo, $32^{\circ} < \ell < 77 ^{\circ}$
and from $168^{\circ} < \ell < 214 ^{\circ}$. This survey uses 7 beam receivers and to date has found, a few tens of new pulsars.

\subsection{RRATs a new type of sources}
These recently discovered RRATs are a new type of radio sources, and are one of the most extreme and powerful radio emitters. 
They were found by \cite{McLaughlin}, in the Parkes Multibeam Survey.
The 11 known RRATs are located at low Galactic latitudes ($|b| < 2^{\circ}$), 
they show burst of radio--emission that last between 2 and 30 ms.
They can not be detected using the standard Fourier techniques used for pulsars making their detection very difficult. 
These sources have periods within the range 0.4--7 s, measured by analysing the arrival times of the individual pulses.
For three of them, there are period derivatives now known,
with values that vary from 7.87 to 12.6 $\times$10$^{-15}$ss$^{-1}$.
These values imply the presence of a very strong magnetic
field ($B \geq \times 10 ^{13}$ G). The true nature of the RRATs is still unknown.
There is no consensus about the origin of their powerful bursts and also about their evolutionary stage. 
They have been related to different
types of neutron stars:  (a) radio quiet X--ray NS population somehow 
related with the AXPs and SGR, possible magnetar candidates (McLaughlin et al. 2006), see also~(\cite{Haberl}) and~(\cite{WT}); 
(b) isolated neutron stars~(\cite{Reynolds});
 (c) re-activated radio pulsars ~(\cite{ZD}) and,
(d) bright pulses of distant pulsars like PSR B0656+14 ~(\cite{Weltevrede}).

Finding new pulsars and new RRAT--like sources will allow us to better understand the life-cycle of massive stars, 
their population, and to test, with better
statistics, the theoretical models that predict the evolution and behaviour of these systems. 
Finally pulsars can provide a uniqe oportunity to study extreme gravitational fields
and are excelent systems for testing General Relativity.

\section{A low frequency survey}

Most of the recent pulsar surveys have been carried out at frequencies near 400 MHz, or at 1.4 GHz for the Parkes Multibeam Survey (\cite{Manch}) and
P--Alfa surveys (\cite{Cord}). Our survey is performed at 328 MHz to maximise the field-of-view/sensitivity trade off. 
The WRST in the 8gr8 mode allows us to explore this frequency range with  better sensitivity than other radio telescopes.
The observations are made using 12 WSRT telescopes which are arranged to form a grating array (i.e. they are equally spaced). The data is then combined in
such a way that we have 8 separate beams pointing at different locations in the primary beam. Thus we get the sensitivity
of all 12 dishes but a beam size of just one! To reach our sensitivity goals and to optimize the data reduction, our observations have
$2^{22}$ samples with a sampling time of 819.2 $\mu$s and a total dwell time of 6872s. In total we have $\sim$ 72 observation points and between
900 and 2200 beams per pointing. For each pointing we have a minimum of two observations, the original observation and a comparison observation. 

\begin{figure}
\centerline{\psfig{file=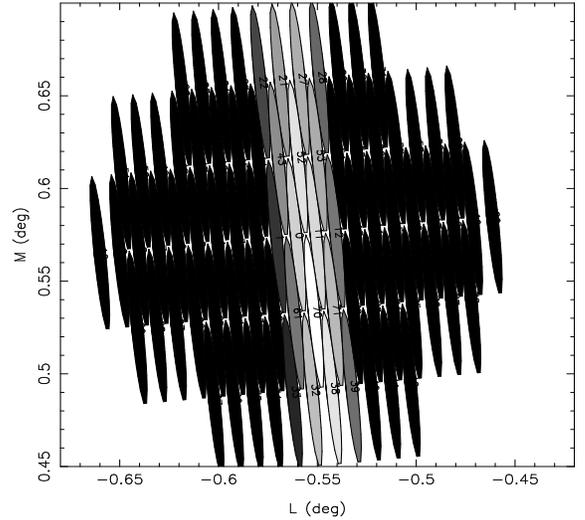,height=8.0cm,width=8.0cm,angle=-90} }
\caption{Plot showing the contour and shape of the beams. The grayscale shows the beam with the largest S/N response for the candidate shown in Fig. 3.
\label{image}}
\end{figure}

\begin{figure*}
\centerline{\psfig{file=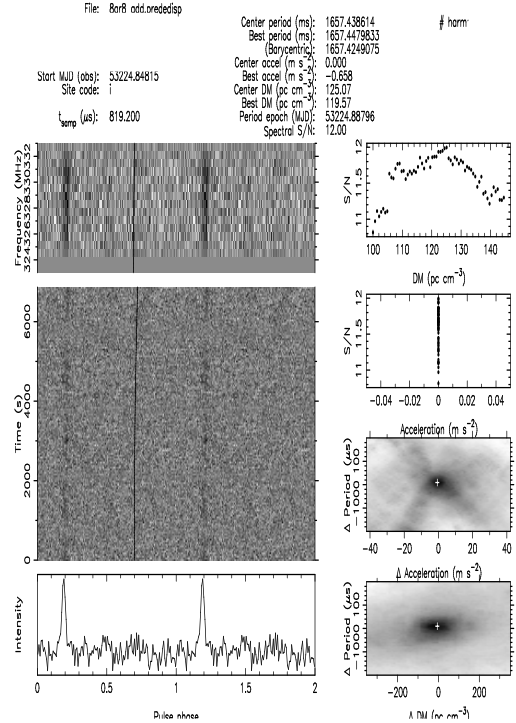,height=7.0cm,width=10cm,angle=-90} }
\caption{Example of the output from a pulsar search of data, a new pulsar candidate from the 8gr8 Cygnus Survey. 
The plots clearly show a pulsating signal and its properties. 
The upper left plot shows a grey--scale plot of frequency vs. pulse phase and shows clearly vertical features that reveal the presence of a pulsating source. 
In the second box on the left side we plot the time series vs. pulse phase, and again we can see the same vertical features. 
Finally the pulsating nature of the source appears clearly in the third box on the left side, where we plot the intensity vs. pulse phase. 
The boxes on the right side show the best DM value for that signal and the distribution of acceleration trials vs period (third box, right) and 
period trials vs DM (lower right box).
\label{image}}
\end{figure*}

\begin{figure*}
\centerline{\psfig{file=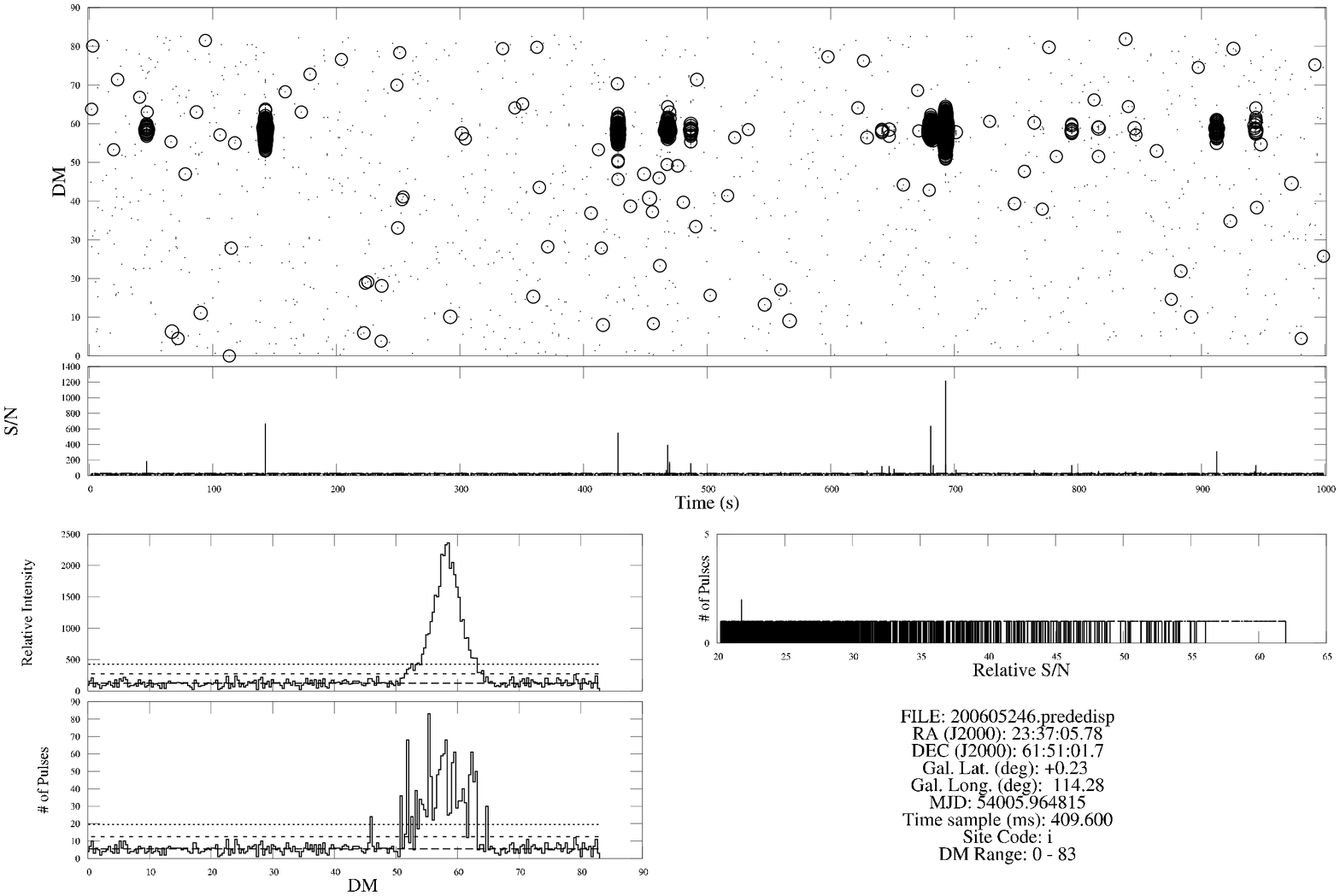,height=8cm,width=14cm,angle=-90} }
\caption{Sample search code for RRATs type objects. The plots correspond to the pulsar PSR B2334+61 and are an example of how a signal (as for RRATs),
should appear.
The upper plot shows the detected signals above a 5--$\sigma$ threshold plotted as circles with a radius
proportional to the SN. The abscisa of both upper boxes shows the arrival time of the signals while the ordinate shows the value of the DM and the value for the 
intensity for the upper and lower box respectivelly. The later one clearly reveals the presence of bright pulses. 
The two lower boxes on the left side show the DM value on the abscisas
while the relative SN on the ordinates and the number of detections They clearly shows a bright source with a high number of counts at DM=58 pc cm$^{-3}$.
Finally the right box below shows the number of detections
with the same SN. Due to the finite width of the bursts coming from the pulsar, many of them are detected at multiple DM values resulting in a vertical broadening
of the features.  Burst which are strongest at zero DM are due to terrestrial interference and are not shown here.
\label{image}}
\end{figure*}

\section{Reducing the data}
The observations were made between 2004 and 2005. Each observation consists of 8 sets of data of 10 MHz wide, per main beam.
The data goes from the receivers located in the focal plane of each antenna 
to a tied array beam former that compensates for delays in arrival times due to the separation of the antenas. 
The data is then digitized using a digital filterbank 
known as the Pulsar Machine--PuMa~(\cite{Voute}) which converts the analogue data into spectra of 512 frequency channels.
These sets of data are stored locally for latter off--line analysis. 

The first step in our analysis is to form dedispersed timeseries for a number of trial DM values.
This is done using a cluster of computers with 31 nodes.
In the same cluster, once the data is dedispersed, we
combine the 8 beams into an array of $N$ sub--beams; the number depends on the position of the main
beam on the sky (see Fig. 2). Each of the $N$ sub-beams is searched for periodicities and for single pulses.
 A typical search run takes between one and two days of computing time.
For our reduction and also for the pulsar search described below, we are using the pse software developed by
Russell Edwards with the single pulse search extensions discussed below.

\subsection{Pulsar search}
For the pulsar search, we use the so--called standard search, calculating an FFT for each time series per DM value.
For the 8gr8 survey our DM range goes from 0 to 1200 pc cm$^{-3}$.
Once a candidate is identified a refined analysis is performed to get the best period and DM value for it. This analysis
includes trying values for linear acceleration steps. The correction for acceleration should lead to a sharper pulse profile for
real binaries.

A possible candidate for a new pulsar should have a plot like that shown in Fig. 3. and also
it must appear at the same position in both of the beam plots for the first and for the comparison observations. In Fig 2 we show one of the beams for
the pulsar candidate shown in Fig 3.

\subsection{Single pulse search}
For the detection of RRATs, we have implemented an approach similar to the one used by ~\cite{McLaughlin}, this
consists of a search in the data for high S/N events with the same DM value as shown in Fig. 4.
We have also implemented a collection of scripts to search for high S$/$N events (above the 5--$\sigma$ threshold
detection limit), and also for events at the same DM values. This will allow us to detect RRAT--like sources and potentially dim pulsars,
without having to manually view many thousands of plots.

\section{Work in progress \& future plans}

The aim of the survey is to locate as many candidates for pulsars and RRATs within the Cygnus region.
We have already found a few candidates for new pulsars and we are already implementing and performing the analysis to
search for RRATs. Besides this, we expect to make a comprehensive analysis of the data we have so far: 

\begin{itemize}
\item[$\bullet$]  We expect to make an individual analysis of sources in order to establish their properties.
\item[$\bullet$] We have completed the search for new pulsars in a large fraction of the 72 observation points in the DM range 
0-700 pc cm$^{-3}$.  We expect to have all the observations reduced in the near future.

\item [$\bullet$] We are currently undertaking the analysis of the 72 observations to search for RRAT sources.

\item [$\bullet$] A long term goal is to perform a multiwavelength
 follow up analysis of the objects to search for possible optical and high energy counterparts.

\end{itemize}

\vspace{6mm}
\begin{acknowledgements}
We gratefully acknowledge the support by the LKBF and UvA for attending this conference.
\end{acknowledgements}

              \clearpage

\end{document}